\documentclass[%
 aip,
 jmp,%
 amsmath,amssymb,
 reprint,%
]{revtex4-1}
\usepackage{color}
\usepackage{graphicx}
\usepackage{dcolumn}
\usepackage{bm}
\usepackage[top=0.6in,bottom=0.6in, left=0.76in, right=0.5in]{geometry}
\begin{document}

\preprint{AIP/123-QED}

\title{Giant magnetothermal conductivity and magnetostriction effect in charge ordered Nd$_{0.8}$Na$_{0.2}$MnO$_{3}$ compound}

\author{B. Samantaray,$^{1}$ N. Khan,$^{1}$ A. Midya,$^{1}$ P. Mandal}
\email{prabhat.mandal@saha.ac.in}
 \affiliation{Saha Institute of Nuclear Physics, 1/AF Bidhannagar, Calcutta 700 064, India.}
 
\author{S. Ravi}
\affiliation{
Department of Physics, Indian Institute of Technology Guwahati, Guwahati - 781039, India}%

\date{\today}

\begin{abstract}
We present results on resistivity ($\rho$), magnetization ($M$), thermal conductivity ($\kappa$), magnetostriction ($\frac{\Delta L}{L(0)}$) and specific heat ($C_{p}$)  of  charge-orbital ordered antiferromagnetic Nd$_{0.8}$Na$_{0.2}$MnO$_{3}$ compound. Magnetic field-induced antiferromagnetic/charge-orbital ordered insulating to ferromagnetic metallic transition leads to giant magnetothermal conductivity  and magnetostriction effect. The low-temperature  irreversibility behavior in $\rho$, $M$, $\kappa$ and $\frac{\Delta L}{L(0)}$ due to field cycling together with striking similarity among  the field and temperature dependence of these parameters manifest the presence of strong and complex spin-charge-lattice coupling in this compound. The giant magnetothermal conductivity is attributed mainly to the suppression of  phonon scattering due to the destabilization of spin fluctuations and static/dynamic Jahn-Teller distortion by the application of magnetic field.
\end{abstract}

\keywords{Magnetothermal conductivity, Magnetostriction and Metamagnetic transition}
\maketitle




Magnetic-field-induced transitions in prototype correlated electron systems, where spin, charge, lattice, and orbital degrees of freedom are coupled strongly, have drawn a considerable attention due to their huge impact on physical properties. Such fascinating phenomena are usually observed in rare-earth based manganites close to half doping. Several field stimulated transitions viz.,  structural transition, first order to second order  ferromagnetic (FM) transition and spin-flop metamagnetic transition from antiferromagnetic (AFM)/charge ordered (CO) state to FM state accompanied by insulator to metal transition have been extensively studied in divalent doped manganite.\cite{tokura,Tomioka95,mahi,Millange,Sarkar09}   However, the study on monovalent hole-doped  manganite systems is very much limited.\cite{Samanta11JSNM,Samanta11JAP,Li,Liu,Repaka} The monovalent doping has the advantage of creation of optimum concentration of Mn$^{4+}$ ions with a relatively small level of doping, as compared to its divalent counterpart. Though there are few reports on the electronic transport, the heat transport and its correlation with magnetic and magnetoelastic properties need to be explored. Magnetothermal transport may provide important insights into  carrier localization due to spin-fluctuations and Jahn-Teller (JT) distortion as a result of strong lattice coupling through the orbital degree of freedom.

Recently, we have investigated the magnetic properties of a medium sized $e_{g}$ electron bandwidth CO system Nd$_{1-x}$Na$_{x}$MnO$_{3}$, \cite{Samanta11JSNM} where the quenching of CO phase was achieved at a lower field relative to other narrowband manganites.\cite{tokura,Millange} Repaka \emph{et al}.\cite{Repaka} have observed giant magneto-thermoelectric power in a quite similar compound and suggested a close interplay between magnetization, electrical resistance and thermoelectric power. In order to understand the nature and strength of spin-lattice coupling and JT distortion  and their role on transport properties, we have carried out  temperature and field dependence of resistivity ($\rho$), magnetization ($M$), thermal conductivity ($\kappa$), magnetostriction ($\frac{\Delta L}{L(0)}$) and specific heat ($C_{p}$) measurements on Nd$_{0.8}$Na$_{0.2}$MnO$_{3}$.  Magnetostriction is a sensitive tool to detect the strength of coupling between magnetic order parameter and lattice structure, i.e., the spin-lattice interaction in the system. The observed giant magnetothermal conductivity, large magnetostriction effect, and their irreversibility characteristics associated with the metamagnetic transition are the manifestations of complex and strong interplay among spin, lattice, and orbital degrees of freedom in the present compound.

The details of Nd$_{0.8}$Na$_{0.2}$MnO$_{3}$ polycrystalline sample preparation and its crystal and magnetic structures are described in  earlier reports.\cite{Samanta11JSNM,Samanta11JAP} The magnetization measurements were carried out by using SQUID VSM (Quantum Design). The temperature and magnetic field dependence of electrical resistivity was measured in a cryostat (Cryogenic Ltd.) equipped with a 9 T superconducting magnet. Thermal conductivity  and specific heat were measured by conventional relaxation method using  Physical Property Measurement System (Quantum Design). The longitudinal magnetostriction  was measured by capacitive method using a miniature tilted-plate dilatometer with applied field parallel to the sample's length.

\begin{figure}
\includegraphics[clip,trim=0cm 0cm 0cm 0.5cm,width=0.4\textwidth]{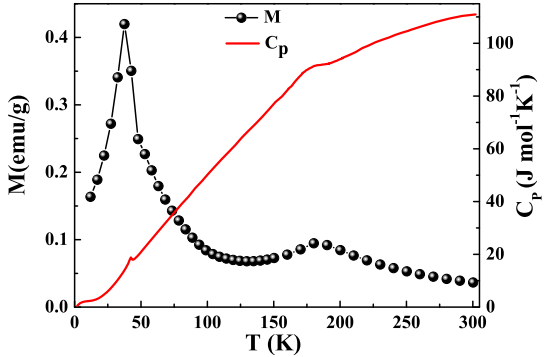}
\caption{Temperature dependence of zero-field-cooled $M$ at 200 Oe and zero-field $C_{P}$ for Nd$_{0.8}$Na$_{0.2}$MnO$_{3}$.}\label{Fig.1}
\end{figure}

The temperature dependence of zero-field-cooled (ZFC) magnetization at 100 Oe as shown in \ref{Fig.1} exhibits a broad maximum at CO temperature, $T_{CO}$=180 K.\cite{Samanta11JAP} With further decreasing temperature, a secondary rise  below 100 K  followed by a cusp-like transition at 42 K is observed in magnetization due to the formation of inhomogeneous/cluster-glass-like magnetic phase. \cite{Samanta11JSNM} The nature of different transitions has been further investigated by measuring the temperature dependence of zero-field specific heat as shown in Fig. 1. The lattice vibrations which contribute towards  $C_{p}$ in terms of phononic excitations are dominant at high temperatures and decrease with decrease in temperature. $C_{p}$ shows a broad peak around $T_{CO}$ and exhibits a weak $\lambda$-like anomaly at 42 K. Similar peak  at $T_{CO}$ is observed in the specific heat data of several half-doped CO manganites.\cite{Lopez} Further decrease in temperature below 10 K reveals a broad anomaly which could be attributed to the Schottky effect due to the splitting of the 4$f$ multiplet of the Nd$^{3+}$ ion.

\begin{figure}
\includegraphics[clip,trim=0cm 0cm 0cm 1cm,width=0.4\textwidth]{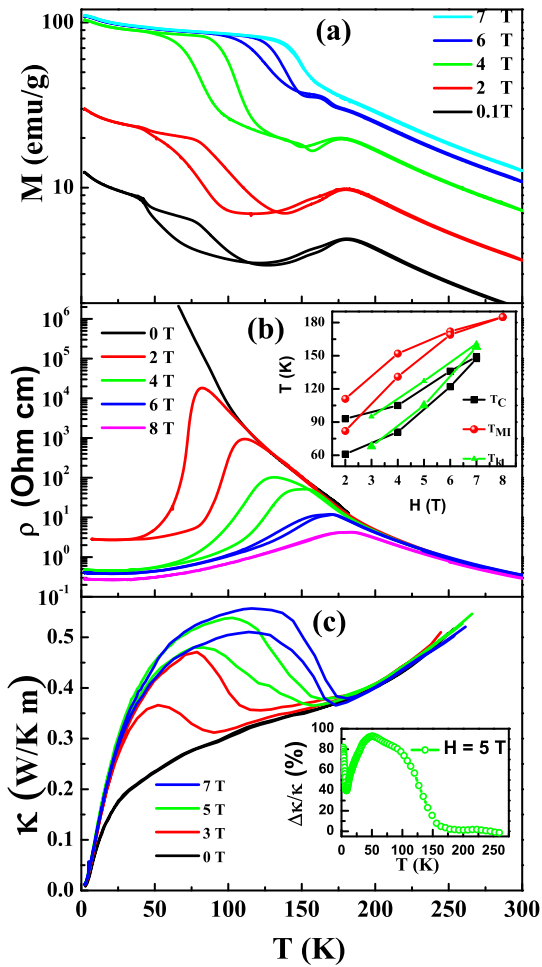}
\caption{(a) $M$ versus $T$, (b) $\rho$ versus $T$ and (c) $\kappa$ versus $T$ plots in FCC and FCW conditions at different magnetic fields. The inset of (b) shows $T_{C}$, $T_{MI}$ and $T_{\kappa}$ in FCC and FCW cycles versus $H$ and inset of (c) shows  $\frac{\Delta \kappa}{\kappa}(\%)$ versus $T$ plots at 5 T.}\label{Fig.2}
\end{figure}
To underscore the nature of interplay between spin, charge and lattice degrees of freedom in the present compound, we have studied magnetization, resistivity, and thermal conductivity as  a function of temperature in both field-cooled-cooling (FCC) and field-cooled-warming (FCW) cycles for fields up to 8 T. Fig. 2(a) shows $M(T)$ at different magnetic fields. With the increase of magnetic field, the value of $M$ increases and the FM transition  ($T_C$) shifts towards higher temperature. This behavior suggests  that the FM phase fraction enhances with magnetic field.  The $\rho(T)$ curves in FCC and FCW conditions for different magnetic fields are shown in Fig. 2(b). In zero field, $\rho$ shows semiconducting behavior and the value of $\rho$ is found to be $\sim$10$^{6}$ $\Omega$ cm at around 65 K. We could not measure $\rho$ below 65 K because of its steep increase.  The application of magnetic field of 2 T suppresses the resistivity by four orders of magnitude at around $T_{MI}$=80  and 111 K in the FCC and FCW cycles, respectively  where the compound undergoes a sharp insulator to metal transition. As the field strength increases, $T_{MI}$ shifts towards higher temperature and the resistivity in the low temperature region decreases substantially. The thermal hysteresis width between FCC and FCW curves is found to be narrowing down progressively with increasing field strength.

The temperature dependence of thermal conductivity at different fields is shown in Fig. 2(c). The zero-field $\kappa$ is  quite small, decreases continuously with decreasing temperature (d$\kappa$/d$T$$>$0) down to 2 K and exhibits a change in slope  below T$_{CO}$. The nature of $T$ dependence of $\kappa$ and its small value  are comparable to that observed in other CO insulator and glassy systems.\cite{Hejtmanek99,Kim,Sharma05} The small value of $\kappa$ could be associated with phonon scattering by magnetic polarons and phononic vibration of Mn$^{3+}$ ions in static and dynamic  JT modes. The phononic contribution to $\kappa$ is mainly determined by the phonon mean free path ($l_{ph}$)  which gets restricted by static and dynamic JT modes.\cite{Visser,Cohn}  Besides this,  the scattering  due to the Nd spin disordering further shortens the phonon mean free path. As a result, $\kappa$ does not show a broad phonon peak owing to the boundary scattering but falls sharply at low temperature. \cite{Hejtmanek99,Sharma04,Wang} For the application of magnetic field of 3 T, $\kappa$ in FCC cycle starts to increase from its zero field value below $\sim$ 120 K and shows a sharp  transition around $T_{\kappa}$ = 96 K.  Similar to $\rho(T)$ and $M(T)$ curves, a large thermal hysteresis is observed between FCC and FCW cycles of $\kappa$($T$). The total thermal conductivity  can be written as sum of phononic ($\kappa_{ph}$), electronic ($\kappa_{e}$), and magnon ($\kappa_{m}$) contributions.  $\kappa_{e}$ at a constant applied magnetic field is estimated  by considering the validation of the Wiedemann-Franz Law, $\kappa_{e}=L_0T/\rho$, where $L_0$ is the Lorentz number. Below $T_{\kappa}$, the deduced value of $\kappa_{e}$ at 5 T is found to be less than 1$\%$ of $\kappa$. So,  $\kappa_{e}$ has negligible contribution towards total thermal conductivity. The magnon thermal conductivity can be estimated from kinetic theory  which is also found to be very small.\cite{Aliev} Though the destabilization of JT distortion by applying magnetic field enhances electronic  contribution to thermal conductivity, the effect of phonon scattering suppression on $\kappa$ is much more significant.  The inset of Fig. 2(c) shows the temperature dependence of magnetothermal-conductivity ($\frac{\Delta\kappa}{\kappa}$), defined as $\frac{\Delta\kappa}{\kappa}$=$\frac{\kappa(H)- \kappa(0)}{\kappa(0)}$. At 5 T, $\frac{\Delta\kappa}{\kappa}$ increases sharply below $T_{CO}$ and exhibits a broad peak. A giant value of $\frac{\Delta\kappa}{\kappa}$ (92$\%$) is observed in the vicinity of cluster-glass transition temperature and $\frac{\Delta\kappa}{\kappa}$ is more than 70$\%$ in the temperature range 27-117 K. The observed value of $\frac{\Delta\kappa}{\kappa}$ is significantly larger as compared to that reported for other doped mangantites. \cite{Cohn,Visser,Chen,Matsukawa} This giant magnetothermal-conductivity suggests that the spin-phonon coupling in the studied sample is quite strong. The transition temperatures, $T_{C}$, $T_{MI}$ and $T_{\kappa}$ in FCC and FCW cycles are obtained from the first derivative  of $M$($T$), $\rho$($T$), and $\kappa$($T$) curves, respectively. The span of the thermal hysteresis at different magnetic fields is estimated from the difference between the transition temperatures in FCC and FCW modes.  The inset of Fig. 2(b) shows that the width of the thermal hysteresis, $\Delta T$, decreases with increasing field and creates a phase separation boundary between CO insulating  and FM metallic regions. Finally, the large thermal hysteresis observed in $M(T)$, $\rho(T)$ and $\kappa(T)$ disappears above 7 T which suggests a change in the nature of transition from first order to second order.

\begin{figure}
\includegraphics[clip,trim=0cm 0cm 0cm 1.5cm,width=0.4\textwidth]{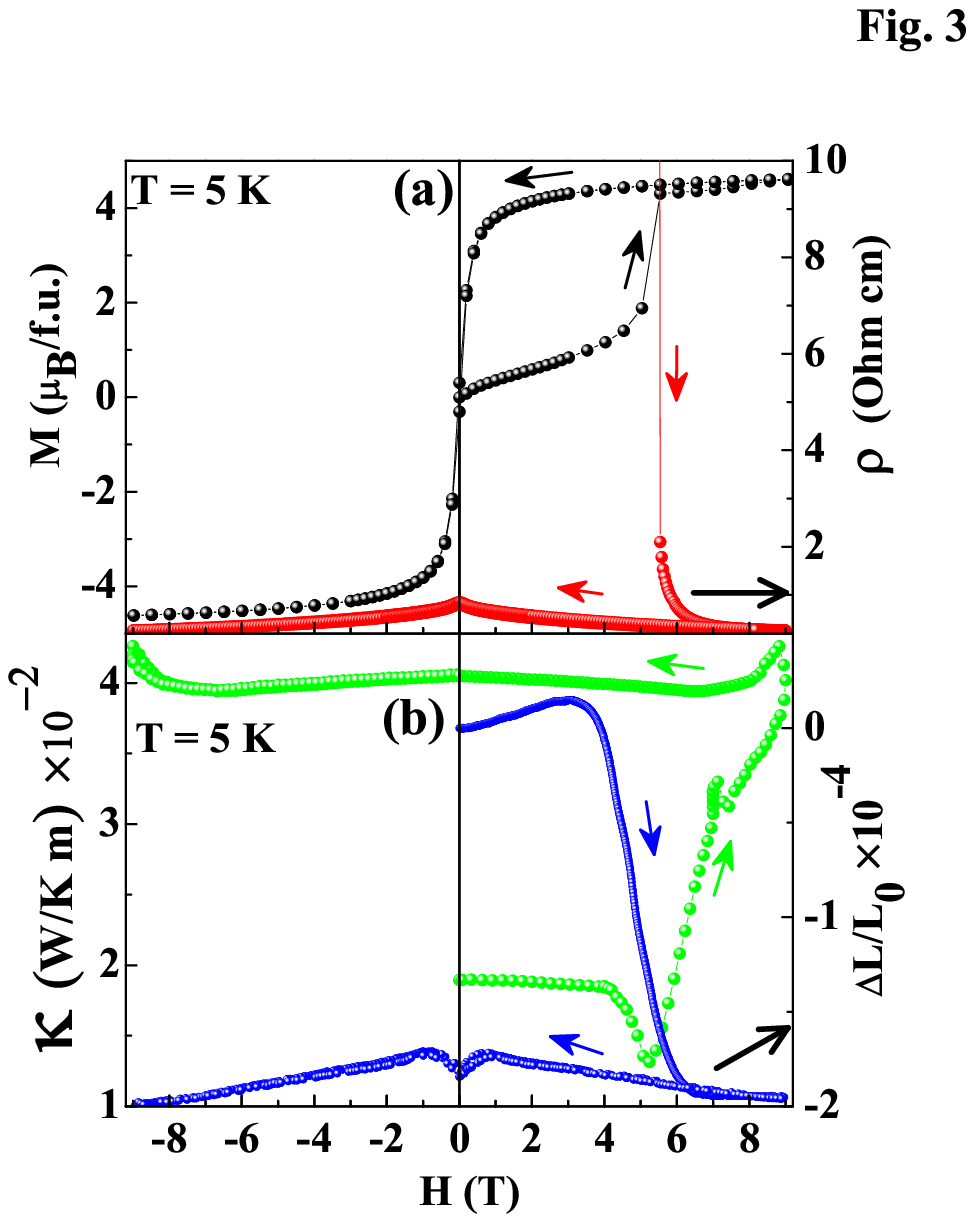}
\caption{Field variation of (a) $M$ and $\rho$, (b) $\kappa$ and $\Delta L/L_{0}$ at 5 K.}\label{Fig.3}
\end{figure}
\begin{figure}
\includegraphics[clip,trim=0cm 0cm 0cm 1.5cm,width=0.4\textwidth]{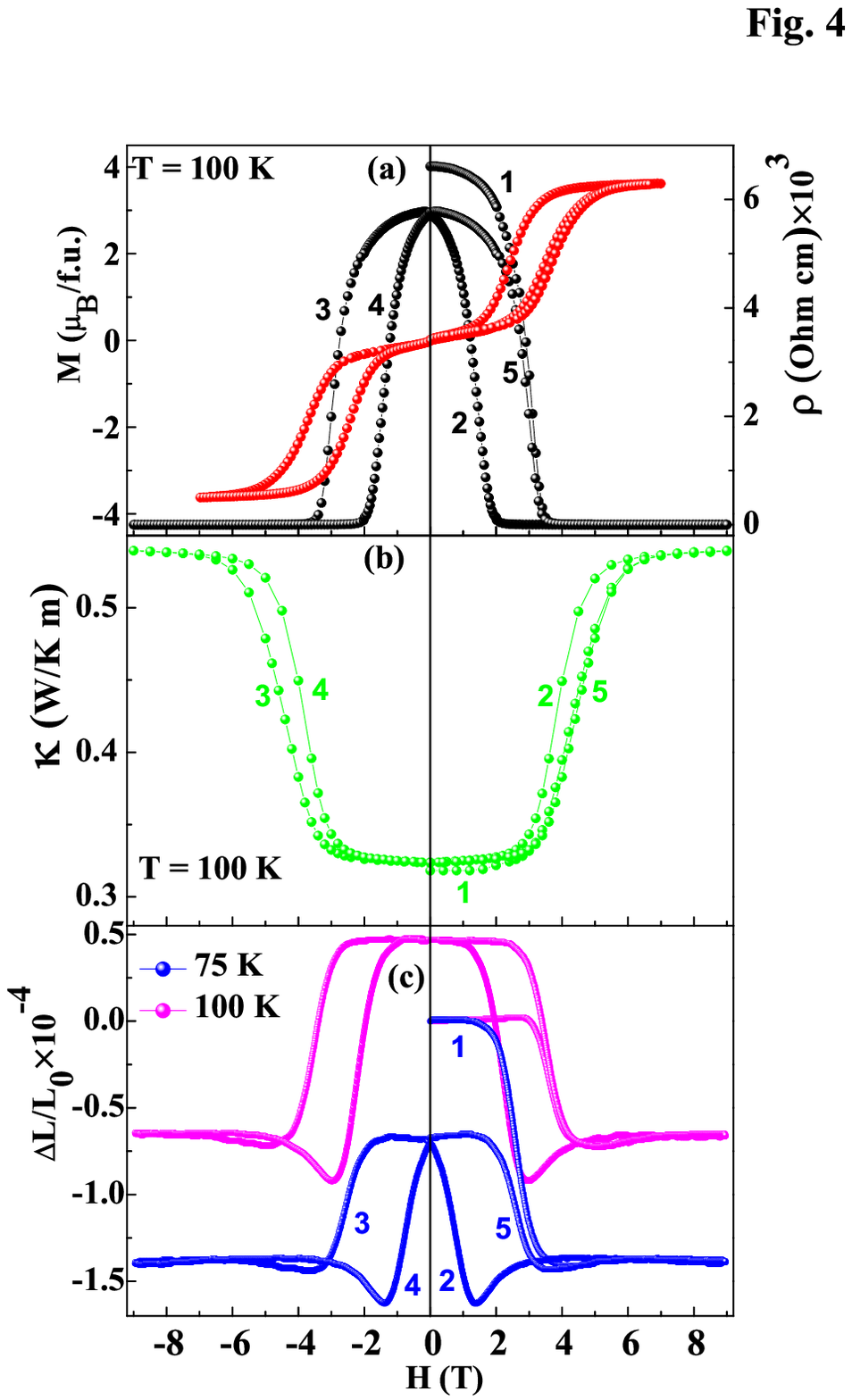}
\caption{Field variation of (a) $M$ and $\rho$ and (b) $\kappa$ at 100 K. (c) Field variation of $\Delta L/L_{0}$ at 75 and 100 K.}\label{Fig.4}
\end{figure}
We have also studied the field dependence of $M$, $\rho$, $\kappa$ and $\frac{\Delta L}{L(0)}$  at different temperatures, where $L(0)$ is the length of the sample at zero-field and $\Delta L$ is the increase in sample length at field $H$. $M(H)$, $\rho(H)$, $\kappa(H)$ and $\frac{\Delta L(H)}{L(0)}$  measured at 5  and 100 K with $\pm$ 9 T field cycling are shown in Figs. 3 and 4, respectively. The $M(H)$ curve at 5 K  [Fig. 3(a)] displays a sharp transition from CO/AFM to FM state above a critical field of $\sim$5 T. Unlike at 5 K, the $M(H)$ loop at 100 K as shown in Fig. 4(a) exhibits a field-induced metamagnetic transition from CO/AFM state to FM state with almost reversible behavior.\cite{Millange} The critical fields for the transition in the ascending and descending branches of the field cycles are 3.5  and 2.5 T, respectively. The $\rho(H)$ curves  show similar irreversible and reversible behavior at 5 K [Fig. 3(a)] and 100 K [Figs. 3(b)], respectively.  At 5 K, a very sharp transition from  highly insulating state to metallic state is observed at a critical field of $\sim$5.5 T in the ascending branch.  Such a sharp transition of $\rho$ from infinitely large value to few $ m \Omega$ cm  also signifies the first-order nature of the field-induced transition. With further field cycling from 9 T to $-$9 T and $-$9 T to 9 T, the system no longer recovers its highly insulating state but  continues to retain the metallic state.   This irreversible behavior is observed up to about 50 K.  The critical fields for the insulating to metallic transition obtained from the ascending branch of the magnetic field for different resistivity isotherms are 2.5, 3.5 and 4.5 T for $T$=75, 100, and 125 K, respectively.

The isotherms of $\kappa$ at 5  and 100 K exhibit very interesting features as can be seen from Figs. 3(b) and 4(b) and show a clear evidence of strong and complex coupling between lattice and spin degrees of freedom. For the virgin curve  at 5 K, $\kappa$  decreases very slowly up to 4 T and then exhibits a dip-like feature at 5 T which is nearly the same field at which the metamagnetic transition occurs [Fig. 3(b)]. With further increasing field above 5 T, $\kappa$ sharply increases as the system undergoes a transition from insulating state to FM metallic state and then exhibits a weak anomaly at 7 T where the secondary rise in the virgin $M(H)$ curve has been observed.  The slow decrease in $\kappa$  up to 4 T is due to the increase of phonon scattering with spin disordering. In an AFM system, when external field is applied, the magnetic moment fluctuation enhances in one of the two AFM sublattices which is antiparallel to $H$. With increasing $H$, more and more spins in the antiparallel sublattice orient along the field direction, which, in turn, increases the spin disordering.  However, the dip-like feature in $\kappa(H)$  can be explained by considering the fact that at very low temperature, the Zeeman energy becomes gapless at some spin-flop field in AFM ordered phase and the energy gap reopens again at high fields.\cite{Wang,Jin} The strongest scattering of phonons by magnons takes place in the vicinity of the spin-flop transition where $\kappa$ attains its lowest value. $\kappa$ shows strong irreversibility behavior for $\pm$9 T field cycling.  On the other hand, $\kappa(H)$ at 100 K as shown in Fig. 4(b) depicts a strong field dependence with sharp change in magnitude in the vicinity of metamagnetic transition at around 3 T but $\kappa(H)$ is found to be almost reversible  in the field cycling process.

In fact, the  longitudinal magnetostriction, $\frac{\Delta L(H)}{L(0)}$,  at different temperatures clearly reflects  the strong spin-lattice coupling and the role of orbital degrees of freedom.  Fig. 3(b) shows $\frac{\Delta L(H)}{L(0)}$ at 5 K and the same at 75, and 100 K is shown in Fig. 4(c).  $\frac{\Delta L(H)}{L(0)}$ at 5 K initially increases  slowly  and then decreases rapidly and changes sign  in the vicinity of CO/AFM to FM transition with increasing field strength. The magnetostriction remains large and negative  during the field cycling process and shows similar irreversibility behavior as observed in $M(H)$, $\rho(H)$, and $\kappa(H)$ at 5 K.   The critical fields for the field-induced transitions obtained from the isothermal  magnetostriction measurements are  5, 4.5, 3 and 2 T for 5, 10, 20 and 50 K, respectively. At 75 K,  $\frac{\Delta L(H)}{L(0)}$ remains almost constant up to a critical field of 1.5 T and  then decreases sharply to negative value.  In the subsequent $\pm$9 T field cycling process,  $\frac{\Delta L(H)}{L(0)}$ exhibits partially reversible behavior. However,  it becomes almost reversible but exhibits both negative and positive magnetostriction effects during the field cycled process at  100 K. The above mentioned sign change in magnetostriction along with field hysteresis are clearly visible up to 125 K and the span of the field hysteresis shrinks with increase in temperature. $\frac{\Delta L(H)}{L(0)}$  is found to be very small, positive and exhibits almost linear behavior for temperatures above $T_{CO}$.

The irreversibility in $M(H)$, $\rho(H)$,  $\kappa(H)$ and $\frac{\Delta L(H)}{L(0)}$ is observed  up to about 50 K and can be explained by kinetic arrest of the high field FM metallic phase. \cite{Chadd} However, above 50 K, the traversed path of the virgin curve falls in the metastable region  which is outside the kinetic arrest band and hence, the characteristic is reversible during the field cycled process due to the de-arrest of the FM metallic phase.\cite{Chadd} The striking similarity in the behavior of $T$ and $H$ dependence of magnetization, charge transport, heat transport and magnetostriction firmly establishes that Nd$_{0.8}$Na$_{0.2}$MnO$_{3}$  is a classic example of spin-charge-lattice coupled system.  Also, the irreversibility in $\kappa(H)$ and $\frac{\Delta L(H)}{L(0)}$  strongly suggests the presence of inhomogeneous long-range strain in the present system.\cite{Sharma05}  The observed features in $\frac{\Delta L(H)}{L(0)}$ can be explained on the basis of real space ordering of Mn$^{3+}$/Mn$^{4+}$ ions accompanied by ${d_{3{x^2} - {r^2}}}/{d_{3{y^2} - {r^2}}}$ orbital ordering of Mn$^{3+}$. Magnetic field stabilizes the ${d_{3{z^2} - {r^2}}}$ orbital state, which results in a destruction of charge ordering and growth of the FM state with spin directed along the $c$ axis.\cite{Kimura} As a result, the MnO$_{6}$ octahedron elongates along $c$ axis and shrinks within $ab$ plane which  brings a net negative magnetostriction above a critical field. At 5 K, we have estimated about 0.07$\%$ volume contraction due to field-induced transition from CO/AFM to FM state.  This large volume contraction is comparable to that observed in Pr$_{0.7}$Ca$_{0.3}$MnO$_{3}$ compound.\cite{garcia} On the contrary, the nature of magneto-volume change in divalent-doped Nd$_{0.5}$Sr$_{0.5}$MnO$_{3}$ is very different where a large volume expansion occurs during the field-induced transition from CO-AFM insulator to FM metallic state. \cite{mahi,Millange}

The magnetic, transport and structural properties and their correlation  have been investigated in CO/AFM Nd$_{0.8}$Na$_{0.2}$MnO$_{3}$ compound.
The large magneto-thermal conductivity is attributed to the suppression of phonon scattering by spin fluctuations and dynamic Jahn-Teller distortion with magnetic field. The giant magneto-thermal conductivity and  magnetostriction associated with field-induced magnetic transition clearly demonstrate  strong spin-phonon coupling in the present compound.

\end{document}